\documentstyle[twoside,epsf]{article}

\catcode`\@=11
\long\def\@makefntext#1{
\protect\noindent \hbox to 3.2pt {\hskip-.9pt  
$^{{\eightrm\@thefnmark}}$\hfil}#1\hfill}		

\def\@makefnmark{\hbox to 0pt{$^{\@thefnmark}$\hss}}	
	
\def\ps@myheadings{\let\@mkboth\@gobbletwo
\def\@oddhead{\hbox{}
\rightmark\hfil\eightrm\thepage}   
\def\@oddfoot{}\def\@evenhead{\eightrm\thepage\hfil
\leftmark\hbox{}}\def\@evenfoot{}
\def\sectionmark##1{}\def\subsectionmark##1{}}

\oddsidemargin=\evensidemargin
\newcounter{sectionc}\newcounter{subsectionc}\newcounter{subsubsectionc}
\renewcommand{\section}[1] {\vspace{12pt}\addtocounter{sectionc}{1} 
\setcounter{subsectionc}{0}\setcounter{subsubsectionc}{0}\noindent 
	{\tenbf\thesectionc. #1}\par\vspace{5pt}}
\renewcommand{\subsection}[1] {\vspace{12pt}\addtocounter{subsectionc}{1} 
\setcounter{subsubsectionc}{0}\noindent 
{\bf\thesectionc.\thesubsectionc. {\kern1pt \bfit #1}}\par\vspace{5pt}}
\renewcommand{\subsubsection}[1] {\vspace{12pt}\addtocounter{subsubsectionc}{1}
	\noindent{\tenrm\thesectionc.\thesubsectionc.\thesubsubsectionc.
	{\kern1pt \tenit #1}}\par\vspace{5pt}}
\newcommand{\nonumsection}[1] {\vspace{12pt}\noindent{\tenbf #1}
	\par\vspace{5pt}}

\newcounter{appendixc}
\newcounter{subappendixc}[appendixc]
\newcounter{subsubappendixc}[subappendixc]
\renewcommand{\thesubappendixc}{\Alph{appendixc}.\arabic{subappendixc}}
\renewcommand{\thesubsubappendixc}
	{\Alph{appendixc}.\arabic{subappendixc}.\arabic{subsubappendixc}}

\renewcommand{\appendix}[1] {\vspace{12pt}
        \refstepcounter{appendixc}
        \setcounter{figure}{0}
        \setcounter{table}{0}
        \setcounter{lemma}{0}
        \setcounter{theorem}{0}
        \setcounter{corollary}{0}
        \setcounter{definition}{0}
        \setcounter{equation}{0}
        \renewcommand{\thefigure}{\Alph{appendixc}.\arabic{figure}}
        \renewcommand{\thetable}{\Alph{appendixc}.\arabic{table}}
        \renewcommand{\theappendixc}{\Alph{appendixc}}
        \renewcommand{\thelemma}{\Alph{appendixc}.\arabic{lemma}}
        \renewcommand{\thetheorem}{\Alph{appendixc}.\arabic{theorem}}
        \renewcommand{\thedefinition}{\Alph{appendixc}.\arabic{definition}}
        \renewcommand{\thecorollary}{\Alph{appendixc}.\arabic{corollary}}
        \renewcommand{\theequation}{\Alph{appendixc}.\arabic{equation}}
        \noindent{\tenbf Appendix \theappendixc #1}\par\vspace{5pt}}
\newcommand{\subappendix}[1] {\vspace{12pt}
        \refstepcounter{subappendixc}
        \noindent{\bf Appendix \thesubappendixc. {\kern1pt \bfit #1}}
	\par\vspace{5pt}}
\newcommand{\subsubappendix}[1] {\vspace{12pt}
        \refstepcounter{subsubappendixc}
        \noindent{\rm Appendix \thesubsubappendixc. {\kern1pt \tenit #1}}
	\par\vspace{5pt}}

\newcommand{\textlineskip}{\baselineskip=13pt}
\newcommand{\smalllineskip}{\baselineskip=10pt}

\newcommand{\publisher}[2]{{\begin{center}\footnotesize\smalllineskip 
	Received #1\\
	Revised #2
	\end{center}
	}}

\def\abstracts#1#2#3{{
	\centering{\begin{minipage}{4.5in}\footnotesize\baselineskip=10pt
	\parindent=0pt #1\par 
	\parindent=15pt #2\par
	\parindent=15pt #3
	\end{minipage}}\par}} 

\def\keywords#1{{
	\centering{\begin{minipage}{4.5in}\footnotesize\baselineskip=10pt
	{\footnotesize\it Keywords}\/: #1
	 \end{minipage}}\par}}
\def\communicate#1{{
	\centering{\begin{minipage}{4.5in}\footnotesize\baselineskip=10pt
	{\footnotesize\it Communicated by}\/: #1
	 \end{minipage}}\par}}

\renewenvironment{thebibliography}[1]
        {\frenchspacing
	 \ninerm\baselineskip=11pt
         \begin{list}{\arabic{enumi}.}
        {\usecounter{enumi}\setlength{\parsep}{0pt}     
	 \setlength{\leftmargin 12.7pt}{\rightmargin 0pt}
         \setlength{\itemsep}{0pt} \settowidth
	{\labelwidth}{#1.}\sloppy}}{\end{list}}

\newcounter{itemlistc}
\newcounter{romanlistc}
\newcounter{alphlistc}
\newcounter{arabiclistc}

\newcommand{\fcaption}[1]{
        \refstepcounter{figure}
        \setbox\@tempboxa = \hbox{\footnotesize Fig.~\thefigure. #1}
        \ifdim \wd\@tempboxa > 5in
           {\begin{center}
        \parbox{5in}{\footnotesize\smalllineskip Fig.~\thefigure. #1}
            \end{center}}
        \else
             {\begin{center}
             {\footnotesize Fig.~\thefigure. #1}
              \end{center}}
        \fi}

\newcommand{\tcaption}[1]{
        \refstepcounter{table}
        \setbox\@tempboxa = \hbox{\footnotesize Table~\thetable. #1}
        \ifdim \wd\@tempboxa > 5in
           {\begin{center}
        \parbox{5in}{\footnotesize\smalllineskip Table~\thetable. #1}
            \end{center}}
        \else
             {\begin{center}
             {\footnotesize Table~\thetable. #1}
              \end{center}}
        \fi}

\def\pmb#1{\setbox0=\hbox{#1}
	\kern-.025em\copy0\kern-\wd0
	\kern.05em\copy0\kern-\wd0
	\kern-.025em\raise.0433em\box0}

\def\fnt#1#2{\footnotetext{\kern-.3em
	{$^{\mbox{\scriptsize #1}}$}{#2}}}

\def\fpage#1{\begingroup
\voffset=.3in
\thispagestyle{empty}\begin{table}[b]\centerline{\footnotesize #1}
	\end{table}\endgroup}

\def\runninghead#1#2{\pagestyle{myheadings}
\markboth{{\protect\footnotesize\it{\quad #1}}\hfill}
{\hfill{\protect\footnotesize\it{#2\quad}}}}
\font\tenrm=cmr10
\font\tenit=cmti10 
\font\tenbf=cmbx10
\font\bfit=cmbxti10 at 10pt
\font\ninerm=cmr9

\font\eightrm=cmr8

\def\FigName{figure}%
\newbox\captionbox
\long\def\@makecaption#1#2{%
  \ifx\FigName\@captype
    \vskip\abovecaptionskip
    \setbox\tempbox\hbox{{\figurecaptionfont #1\hskip1em #2}}
	\ifdim\wd\tempbox< 28pc
	\centerline{\box\tempbox}
	\else
	{\figurecaptionfont #1\hskip1em #2\par}
\fi\else
  	\setbox\tempbox\hbox{{\tablecaptionfont #1\hskip1em #2}}
 	\ifdim\wd\tempbox< 28pc 
	\centerline{\box\tempbox}
	\else
	{\tablecaptionfont #1\hskip1em #2\par}%
	\fi   
 \vskip\belowcaptionskip
 \fi}
\InputIfFileExists{psfig.sty}
{\typeout{^^Jpsfig.sty inputed...ok}}{\typeout{^^JWarning: psfig.sty could be be found.^^J}}
\InputIfFileExists{epsfsafe.tex}
{\typeout{^^Jepsfsafe.tex inputed...ok}}
			{\typeout{^^JWarning: epsfsafe.tex could not be found.^^J}}
\InputIfFileExists{epsfig.sty}
{\typeout{^^Jepsfig.sty inputed...ok}}{\typeout{^^JWarning: epsfig.sty could not be found.^^J}}
\InputIfFileExists{epsf.sty}
{\typeout{^^Jepsf.sty inputed...ok}}{\typeout{^^JWarning: epsf.sty could not be found.^^J}}%
\def\fps@figure{tbp}
\def\ftype@figure{1}
\def\ext@figure{lof}
\def\fnum@figure{Fig.\ \thefigure}
\def\qed{\hbox{${\vcenter{\vbox{	          
   \hrule height 0.4pt\hbox{\vrule width 0.4pt height 6pt
   \kern5pt\vrule width 0.4pt}\hrule height 0.4pt}}}$}}

\begin{document}
\setlength{\textheight}{8.0truein}    

\runninghead{Title   $\ldots$} 
            {Author(s) $\ldots$}

\normalsize\textlineskip
\thispagestyle{empty}
\setcounter{page}{1}


\vspace*{0.88truein}

\fpage{1}
\centerline{\bf
MEASURABLE LOWER BOUNDS ON CONCURRENCE}
\vspace*{0.035truein}
\vspace*{0.37truein}
\centerline{\footnotesize 
IMAN SARGOLZAHI\footnote{sargolzahi@stu-mail.um.ac.ir, sargolzahi@gmail.com} , SAYYED YAHYA MIRAFZALI and MOHSEN SARBISHAEI}
\vspace*{0.015truein}
\centerline{\footnotesize\it Department of Physics, Ferdowsi University of Mashhad
}
\baselineskip=10pt
\centerline{\footnotesize\it Mashhad, Iran}
\vspace*{0.225truein}
\publisher{(received date)}{(revised date)}

\vspace*{0.21truein}
\abstracts{
We derive measurable lower bounds on concurrence of arbitrary mixed states, for both bipartite
and multipartite cases. First, we construct  measurable lower bonds on the \textit{purely algebraic} bounds
 of concurrence [F. Mintert \textit{et al.} (2004), Phys. Rev. lett., 92, 167902]. Then, using the fact that the sum of 
the square of the algebraic bounds is a lower bound of the squared concurrence, we sum over our measurable bounds
 to achieve a measurable lower bound on concurrence. With two typical examples, we show 
 that our method can detect more entangled 
 states and also can give sharper lower bonds than the similar ones.}{}{}

\vspace*{10pt}
\vspace*{3pt}
\communicate{to be filled by the Editorial}

\vspace*{1pt}\textlineskip	
\section{Introduction}	        
\vspace*{-0.5pt}
\noindent
Recently, many studies have been focused on the experimental quantification
of entanglement~\cite{1}. Bell inequalities and entanglement witnesses~\cite{1,2} 
can be used to detect entangled states experimentally, but they don't give any information about the 
amount of entanglement. In addition, quantum state tomography~\cite{3}, determination of the full density 
operator $\rho$ by measuring a complete set of observables, is 
only practical for low dimensional systems since the number of measurements needed 
for it grows rapidly as the dimension of the system increases. Fortunately, several 
methods have been introduced which let one to estimate experimentally the amount of the entanglement of 
an unknown $\rho$ with no need to the full tomography~\cite{1,4,5,6,7,8,9,10,
11,12,13,14,15,15a,16,17,18,19,20,21,22,23,24,27,28,28p,29,29b}. A simple and 
straightforward method is the one introduced in~\cite{8,14,17} for finding measurable lower 
bounds on an entanglement measure, namely the \textit{concurrence}~\cite{30}. These lower bounds are 
in terms of the expectation values of some Hermitian operators with respect to two-fold
or one-fold copy of $ \rho $. It is worth noting that these bounds work well for weakly mixed
states~\cite{31,8,14,17,5}.

In this paper we will use a similar procedure as~\cite{8,14}
to construct measurable lower bounds on the \textit{purely algebraic bounds of concurrence}~\cite{32,30}.
In addition, using a theorem in Sec. II, we  show that  the sum of our measurable bounds leads to a measurable lower bound on the \textit{concurrence} itself.
Then, we show that this method gives better results than those introduced in~\cite{8,14} for two typical examples.  

The paper is organized as follows. In Sec. II, the concurrence and its $ MKB $ (Mintert-Kus-Buchleitner) lower bounds~\cite{32} are introduced. In Secs. III and IV, we propose measurable lower bounds on the purely algebraic bounds of concurrence~\cite{32}, which are a special class of 
$ MKB $ bounds. The generalization to the multipartite case is given in Sec. V and we end
this paper in Sec. VI with a summary and discussion.


\section{Concurrence and its $ MKB $ Lower Bounds}
\noindent
For a pure bipartite state $ \vert\Psi\rangle $, $ \vert\Psi\rangle\in\mathcal{H}_{A}\otimes \mathcal{H}_{B}\ $,
concurrence is defined as~\cite{30}:
\begin{equation}
C\left( \Psi\right)=\sqrt{2 [\langle\Psi\vert\Psi\rangle^{2}-tr\rho_{r}^{2} ]}\,,
\label{1}
\end{equation}
where $ \rho_{r} $ is the reduced density operator obtained by tracing over either   
subsystems A or B. It is obvious  that iff $ \vert\Psi\rangle $ is a product state, i.e.  
$ \vert\Psi\rangle= \vert\Psi_{A}\rangle\otimes \vert\Psi_{B}\rangle  $, then $ C(\Psi)=0 $. 
Interestingly, $ C(\Psi) $ can be written in terms of the expectation value of an 
observable with respect to two identical copies of $ \vert\Psi\rangle $~\cite{30,11,12}: 
\begin{eqnarray}
C\left( \Psi\right)=\sqrt{_{AB}\langle\Psi\vert_{AB}\langle\Psi\vert{\cal A} \vert\Psi\rangle_{AB}\vert\Psi\rangle_{AB}} \,,\cr 
{\cal A}=4 P_{-}^{A}\otimes P_{-}^{B}\,,\qquad\qquad
\label{2}        
\end{eqnarray}
where $ P_{-}^{A} $ ($ P_{-}^{B} $) is the projector onto the antisymmetric subspace of $ \mathcal{H}_{A}\otimes \mathcal{H}_{A}\ $ ($ \mathcal{H}_{B}\otimes \mathcal{H}_{B}\ $).
A possible decomposition of $ {\cal A} $ is 
\begin{eqnarray}
{\cal A}=\sum_{\alpha}\vert\chi_{\alpha}\rangle\langle\chi_{\alpha}\vert \,,\qquad\quad\cr
\vert\chi_{\alpha}\rangle\ =\large(\vert xy\rangle - \vert yx\rangle)_{A}\large(\vert pq\rangle - \vert qp\rangle)_{B}\,,
\label{3}
\end{eqnarray}
where $ \vert x\rangle $ and $ \vert y\rangle $ ($ \vert p\rangle $ and $ \vert q\rangle $) are
two different members of an orthonormal basis of the A (B) subsystem.
For mixed states the concurrence is defined as follows~\cite{30}:
\begin{eqnarray}
C(\rho)= \min \sum_{i}p_{i}C( \Psi_i )\,, \qquad\qquad\cr 
\rho =\sum_{i}p_{i} \vert\Psi_{i}\rangle\langle\Psi_{i}\vert\,, \qquad p_{i}\geq 0\,, \qquad\sum_{i}p_{i} =1\,,
\label{4}
\end{eqnarray}
where the minimum is taken over all decompositions of $ \rho $ into pure states $ \vert\Psi_{i}\rangle $. It is appropriate to write $C(\rho)$ in terms of the subnormalized states $ \vert\psi_{i}\rangle $ rather than the 
normalized ones $\vert\Psi_{i}\rangle $:
\begin{eqnarray}
C(\rho)= \min \sum_{i}\sqrt{\langle\psi_{i}\vert\langle\psi_{i}\vert{\cal A} \vert\psi_{i}\rangle\vert\psi_{i}\rangle}\,,\cr 
\vert\psi_{i}\rangle=\sqrt{p_{i}}\vert\Psi_{i}\rangle\,,\qquad \rho =\sum_{i} \vert\psi_{i}\rangle\langle\psi_{i}\vert \,;
\label{5}
\end{eqnarray}
since all decompositions of $ \rho $ into subnormalized states are related to each other by unitary 
matrices~\cite{3}: consider an arbitrary decomposition of $ \rho =\sum_{j} \vert\varphi_{j}\rangle\langle\varphi_{j}\vert $ (As a special case, one can choose 
 $ \vert\varphi_{j}\rangle=
  \sqrt{\lambda_{j}}\vert\Phi_{j}\rangle $, where $ \vert\Phi_{j}\rangle $ and $ \lambda_{j} $ are eigenvectors and
eigenvalues of $ \rho $ respectively:
  $ \rho =\sum_{j}\lambda_{j} \vert\Phi_{j}\rangle\langle\Phi_{j}\vert $.), for any other decomposition of $ \rho =\sum_{i} \vert\psi_{i}\rangle\langle\psi_{i}\vert $ we have~\cite{3}:
  \begin{equation}
\vert\psi_{i}\rangle =\sum_{j}U_{ij}\vert\varphi_{j}\rangle\,,\qquad \sum_{i} U^{\dagger}_{ki} U_{ij} =
 \delta_{jk}\,.
 \label{6}
\end{equation}
So Eq. (\ref{5}) can be written as:
\begin{eqnarray}
C(\rho)= \min_{U} \sum_{i}\sqrt{\sum_{jklm}U_{ij}U_{ik}{\cal A}^{lm}_{jk}U^{\dagger}_{li}U^{\dagger}_{mi}}\,,\cr 
{\cal A}^{lm}_{jk}= \langle\varphi_{l}\vert\langle\varphi_{m}\vert {\cal A} \vert\varphi_{j}\rangle\vert
\varphi_{k}\rangle\,.\qquad\qquad
\label{7}
\end{eqnarray}

From the definition of $ C(\rho) $ in Eq. (\ref{4}) it is obvious that $ C(\rho)=0 $ iff $ \rho $ can be 
decomposed into product states. In other words, $ C(\rho)=0 $ iff $ \rho $ is separable. In addition, it can be
shown that the concurrence is an entanglement monotone~\cite{33} (An entanglement monotone is a function of 
$ \rho $ which does not increase, on average, under LOCC and vanishes for separable states~\cite{34}.).
But, except for the two-qubit case~\cite{34a}, $ C(\rho) $ can not be computed in general; i.e., in general, one can not find the $ U $ 
which minimizes Eq. (\ref{7}). Any numerical method for finding the $ U $ which minimizes Eq. (\ref{7}) leads to
an upper bound for $ C(\rho) $. So, finding lower bounds on $ C(\rho) $ is desirable. So far, several 
lower bounds for $ C(\rho) $ have been introduced~\cite{32,30,35,36,36a,36b,37,38,39,5,8,13,14,17,19,27,28,28p,29}. One of them is 
that introduced by F. Mintert \textit{et al.} in~\cite{32,30}. Now, we redrive their lower bounds 
in a slightly different form to make them more suitable for finding measurable lower bounds in the following sections.

Assume that the decomposition of $ \rho $ which minimizes Eq. (\ref{5}) is $ \rho =\sum_{j}\vert\xi_{j}\rangle\langle\xi_{j}\vert $, then from Eqs. (3) and (5), we have:   
\begin{eqnarray}
C(\rho)= \sum_{j}\sqrt{\sum_{\alpha}\vert\langle\chi_{\alpha}\vert\xi_{j}\rangle\vert\xi_{j}\rangle\vert^{2}}\geq\sum_{j}
\vert\langle\chi_{\beta}\vert\xi_{j}\rangle\vert\xi_{j}\rangle\vert 
\geq\min_{\left\lbrace \vert\psi_{i}\rangle\right\rbrace }
\sum_{i}\vert\langle\chi_{\beta}\vert\psi_{i}\rangle\vert\psi_{i}\rangle\vert\,,\qquad
\label{8}
\end{eqnarray}
where $ \vert\chi_{\beta}\rangle\in\left\lbrace \vert\chi_{\alpha}\rangle\right\rbrace $, and the minimum is taken over all decompositions of $ \rho $ as $ \rho =\sum_{i}\vert\psi_{i}\rangle\langle\psi_{i}\vert $.
Now, using Eq. (\ref{6}), we have:
\begin{eqnarray}
\min_{\lbrace\vert\psi_{i}\rangle\rbrace} \sum_{i}\vert\langle\chi_{\beta}\vert\psi_{i}\rangle\vert\psi_{i}\rangle\vert 
=\min_{U} \sum_{i}\vert \sum _{jk} U_{ij}T_{jk}^{\beta}U_{ki}^{\top}\vert 
=\min_{U}\sum_{i}\vert\left[ UT^{\beta}U^{\top}\right]_{ii}\vert\, ,\cr 
 T_{jk}^{\beta} =\langle\chi_{\beta}\vert\varphi_{j}\rangle\vert\varphi_{k}\rangle\,.\qquad\qquad\qquad\qquad\qquad\qquad
\label{9}
 \end{eqnarray}
Since $ T^{\beta} $ is a symmetric matrix, the minimum in Eq. (\ref{9}) can be computed and we have~\cite{30}:
\begin{equation}
\min_{U}\sum_{i}\vert\left[ UT^{\beta}U^{\top}\right]_{ii}\vert =\max \lbrace 0,S_{1}^{\beta}-
\sum_{l>1}S_{l}^{\beta}\rbrace\,,
\label{10}  
\end{equation}
where $ S_{l}^{\beta} $ are the singular values of $ T^{\beta} $, in decreasing order. The above expression is what was named 
\textit{purely algebraic lower bound} of concurrence in ~\cite{30,32} and we will refer to it as $ ALB(\rho) $.

Let us define
\begin{equation}
\vert\tau\rangle =\sum_{\alpha}z_{\alpha}^{\ast}\vert\chi_{\alpha}\rangle\,,\qquad \sum_{\alpha}\vert z_{\alpha}\vert^{2}=1\,.\qquad
\label{11}
\end{equation}
Obviously, $ \vert\tau\rangle $ is an element of another (normalized to 2) basis of 
$ P_{-}^{A}\otimes P_{-}^{B} $, $ \lbrace\vert\chi_{\alpha}^{\prime}\rangle\rbrace$. Then:
\begin{eqnarray}
\vert\tau\rangle\equiv\vert\chi_{1}^{\prime}\rangle\,,\qquad\qquad\qquad\cr 
{\cal A} =\sum_{\alpha}\vert\chi_{\alpha}\rangle\langle\chi_{\alpha}\vert =\vert\tau\rangle\langle\tau\vert +
\sum_{\alpha >1}\vert\chi_{\alpha}^{\prime}\rangle\langle\chi_{\alpha}^{\prime}\vert\,. 
\label{12}
\end{eqnarray}  
Again, as the inequality (\ref{8}), we have: 
\begin{eqnarray}
C(\rho)= \sum_{j}\sqrt{\sum_{\alpha}\vert\langle\chi_{\alpha}^{\prime}\vert\xi_{j}\rangle\vert\xi_{j}
\rangle\vert^{2}}\geq\sum_{j}
\vert\langle\tau\vert\xi_{j}\rangle\vert\xi_{j}\rangle\vert\cr 
\geq\min_{\left\lbrace \vert\psi_{i}\rangle\right\rbrace }
\sum_{i}\vert\langle\tau\vert\psi_{i}\rangle\vert\psi_{i}\rangle\vert\qquad\qquad\qquad\qquad \cr =\min_{U}\sum_{i}\vert\left[ U{\cal T} U^{\top}\right]_{ii}\vert  =\max \lbrace 0,S_{1}^{\tau}-
\sum_{l>1}S_{l}^{\tau}\rbrace\,,
\cr 
{\cal T}_{jk} =\langle\tau\vert\varphi_{j}\rangle\vert\varphi_{k}\rangle =\sum_{\alpha}z_{\alpha}T_{jk}^{\alpha}\,,
\qquad\qquad
\label{13}
\end{eqnarray}
where $ S_{l}^{\tau} $ are the singular values of $ {\cal T} $, in decreasing order. The above expression is the general form of the lower bounds introduced in ~\cite{32,30} and we call it $ LB(\rho) $.

We end this section by proving a useful theorem: if $ \left\lbrace \vert\chi_{\alpha}^{\prime}\rangle\right\rbrace  $ be an orthogonal
 (normalized to 2) basis of $ P_{-}^{A}\otimes P_{-}^{B} $, i.e. $ {\cal A} =\sum_{\alpha}\vert\chi_{\alpha}^{\prime}\rangle\langle\chi_{\alpha}^{\prime}\vert$, then:
\begin{eqnarray}
C^{2}(\rho)= \sum_{ij}\sqrt{\sum_{\alpha}\vert\langle\chi_{\alpha}^{\prime}\vert\xi_{i}\rangle\vert\xi_{i}
\rangle\vert^{2}}\sqrt{\sum_{\alpha}\vert\langle\chi_{\alpha}^{\prime}\vert\xi_{j}\rangle\vert\xi_{j}
\rangle\vert^{2}}\cr \geq \sum_{ij}\sum_{\alpha}\vert\langle\chi_{\alpha}^{\prime}\vert\xi_{i}\rangle\vert\xi_{i}
\rangle\vert\vert\langle\chi_{\alpha}^{\prime}\vert\xi_{j}\rangle\vert\xi_{j}
\rangle\vert \qquad\qquad\cr =\sum_{\alpha}\left( \sum_{i}\vert\langle\chi_{\alpha}^{\prime}\vert\xi_{i}\rangle\vert\xi_{i}
\rangle\vert\right) ^{2}\geq\sum_{\alpha}\left[  LB_{\alpha}(\rho)\right] ^{2}\,,\cr 
 LB_{\alpha}(\rho) =\min_{\left\lbrace \vert\psi_{i}\rangle\right\rbrace }
\sum_{i}\vert\langle\chi_{\alpha}^{\prime}\vert\psi_{i}\rangle\vert\psi_{i}\rangle\vert\,.\qquad
\label{14}
\end{eqnarray}
In proving the above theorem we have used the Cauchy-Schwarz inequality. Obviously, any entangled $ \rho $
which can not be detected by $ {LB_{\alpha}} $, can not be detected by $ \sum_{\alpha}\left[  LB_{\alpha}(\rho)\right] ^{2} $ either; i.e.,
 $ \sum_{\alpha}\left[  LB_{\alpha}(\rho)\right] ^{2} $ is not
a more powerful criteria than $ {LB_{\alpha}} $, but, quantitatively, it may lead to a better lower bound for
$ C(\rho) $.

It should be mentioned that the above theorem is, in fact, the generalization of what has been proved in ~\cite{38}.
There, it was shown that:
\begin{eqnarray}
\tau(\rho) =\sum C_{mn}^{2}(\rho)\leq C^{2}(\rho)\,,\qquad\qquad\cr 
 C_{mn}(\rho) =\min_{\left\lbrace \vert\psi_{i}\rangle\right\rbrace }
\sum_{i}\vert\langle\psi_{i}\vert L_{m_{A}}\otimes L_{n_{B}}\vert\psi_{i}^{\ast}\rangle\vert\,,
\label{15} 
\end{eqnarray}
where $ L_{m_{A}} $ and $ L_{n_{B}} $ are generators of $ SO(d_{A}) $ and $ SO(d_{B})$ respectively 
$ (d_{A/B} = dim(\mathcal{H}_{A/B})) $, and $ \vert\psi_{i}^{\ast}\rangle $ is the complex conjugate of
$ \vert\psi_{i}\rangle $ in the computational basis. In this basis $ L_{m_{A}} $ and $ L_{n_{B}} $ are~\cite{40}:
\begin{eqnarray*}
L_{m_{A}} = \vert x\rangle_{A}\langle y\vert -\vert y\rangle_{A}\langle x\vert\,,\qquad\quad 
L_{m_{B}} = \vert p\rangle_{B}\langle q\vert -\vert q\rangle_{B}\langle p\vert\,.
\end{eqnarray*}
For an arbitrary $ \vert\psi\rangle $, according to the definition of $\vert\chi_{\alpha}\rangle$ in 
Eq. (\ref{3}), it can be seen that:
\begin{equation}
\vert\langle\psi\vert L_{m_{A}}\otimes L_{n_{B}}\vert\psi^{\ast}\rangle\vert =
\vert\langle\chi_{\alpha}\vert\psi\rangle\vert\psi\rangle\vert\,.
\label{16}
\end{equation}
So:
\begin{eqnarray}
C_{mn}(\rho)= ALB_{\alpha}(\rho)\,,\qquad\qquad\cr 
ALB_{\alpha}(\rho)=\min_{\left\lbrace \vert\psi_{i}\rangle\right\rbrace }
\sum_{i}\vert\langle\chi_{\alpha}\vert\psi_{i}\rangle\vert\psi_{i}\rangle\vert\,.
\label{17}
\end{eqnarray}
So what was proved in~\cite{38} is, in fact, the special case of 
$\vert\chi_{\alpha}^{\prime}\rangle =\vert\chi_{\alpha}\rangle $ in expression (\ref{14}). In addition, since 
$ ALB_{\alpha} $ can detect bound entangled states ~\cite{32,30}, this claim of~\cite{38} that any state for 
which $ \tau(\rho)>0 $ is distillable, is not correct.


\section{Measurable Lower Bounds in terms of Two Identical Copies of $ \rho $}
\noindent
As we have seen in Eq. (\ref{2}) the concurrence of a pure state $\vert\Psi\rangle $ can be written in terms of the 
expectation value of the observable $ {\cal A} $ with respect to two identical copies of $\vert\Psi\rangle $. For an arbitrary 
mixed state $ \rho_{AB} $, it has been shown that~\cite{8}:
\begin{eqnarray}
C^{2}(\rho_{AB})\geq tr\left( \rho_{AB}\otimes \rho_{AB} V_{(i)}\right)\,, \qquad i=1,2\,;\qquad\cr 
V_{(1)} =4 \left(P_{-}^{A}-P_{+}^{A}\right) \otimes P_{-}^{B}\,, \qquad\qquad
V_{(2)} =4 P_{-}^{A}\otimes\left(P_{-}^{B}-P_{+}^{B}\right)\,,
\label{18}
\end{eqnarray}
where $ P_{+}^{A} $ ($ P_{+}^{B} $) is the projector onto the symmetric subspace of $ \mathcal{H}_{A}\otimes \mathcal{H}_{A}\ $ ($ \mathcal{H}_{B}\otimes \mathcal{H}_{B}\ $). The above expression means that measuring $ V_{(i)} $ on two identical copies of $ \rho $, i.e.  
$ \rho\otimes\rho $, gives us a measurable \textit{lower} bound on $ C^{2}(\rho) $. 
It is worth noting that if the entanglement of $\rho$ can be detected by $ V_{(i)} $, then $\rho$ is 
distillable~\cite{29}. 

As one can see from expression (\ref{13}), the $LB$ of a pure state $\vert\Psi\rangle$ can also be written in terms of the expectation value of the observable $\vert\tau\rangle\langle\tau\vert$ with respect to two identical copies of $\vert\Psi\rangle$. Now, for an arbitrary mixed state $ \rho $, can we find an observable $V$ such that the following inequality holds?
\begin{equation}
LB^{2}(\rho)\geq tr\left( \rho\otimes\rho V\right)\,,
\label{19} 
\end{equation}
Fortunately for the special case of  
$ \vert\tau\rangle =\vert\chi_{\alpha}\rangle$, where $\vert\chi_{\alpha}\rangle$ are defined in Eq. (\ref{3}), we can do so.

Assume that the decomposition of $\rho$ which gives the minimum in Eq. (\ref{9}) is $\rho =\sum_{i}\vert\theta_{i}^{\alpha}\rangle\langle\theta_{i}^{\alpha}\vert$, i.e.:
\begin{equation}
ALB_{\alpha}(\rho) = \sum_{i}\vert\langle\chi_{\alpha}\vert\theta_{i}^{\alpha}\rangle\vert\theta_{i}^{\alpha}\rangle\vert\,.
\label{20}
\end{equation}
In addition, assume that for a Hermitian operator $V_{\alpha}$, which acts on 
$ \mathcal{H}_{A}\otimes \mathcal{H}_{B}\otimes \mathcal{H}_{A}\otimes \mathcal{H}_{B}\ $, and arbitrary 
$ \vert\psi\rangle$ and $\vert\varphi\rangle$, $\vert\psi\rangle\in \mathcal{H}_{A}\otimes \mathcal{H}_{B} $ and
$\vert\varphi\rangle\in \mathcal{H}_{A}\otimes \mathcal{H}_{B} $, we have:
\begin{equation}
\vert\langle\chi_{\alpha}\vert\psi\rangle\vert\psi\rangle\vert\vert\langle\chi_{\alpha}\vert\varphi\rangle
\vert\varphi\rangle\vert\geq\langle\psi\vert\langle\varphi\vert V_{\alpha}\vert\psi\rangle\vert\varphi\rangle\,.
\label{21}
\end{equation}
Now, from the expressions (20) and (21), we have:
\begin{eqnarray}
ALB_{\alpha}^{2}(\rho) = \sum_{ij}\vert\langle\chi_{\alpha}\vert\theta_{i}^{\alpha}\rangle\vert\theta_{i}^{\alpha}\rangle\vert
\vert\langle\chi_{\alpha}\vert\theta_{j}^{\alpha}\rangle\vert\theta_{j}^{\alpha}\rangle\vert 
\geq\sum_{ij}\langle\theta_{i}^{\alpha}\vert\langle\theta_{j}^{\alpha}\vert V_{\alpha}\vert\theta_{i}^{\alpha}\rangle\vert\theta_{j}^{\alpha}\rangle =tr \left( \rho\otimes\rho V_{\alpha}\right)\,.
\label{22} 
\end{eqnarray}
So, for any $ V_{\alpha}$ satisfying inequality (21), measuring $ V_{\alpha}$ on two identical copies of $\rho$ 
gives a lower bound on $ALB_{\alpha}^{2}(\rho)$. We can prove that the inequality (21) holds for(see the Appendix): 
\begin{eqnarray}
V_{\alpha} =V_{(1)\alpha}={\cal M}V_{(1)} {\cal M}\,,\quad\qquad
V_{\alpha} =V_{(2)\alpha}={\cal M}V_{(2)} {\cal M}\,,\cr 
{\cal M}={\cal M}_{A}\otimes{\cal M}_{A}\otimes{\cal M}_{B}\otimes{\cal M}_{B}\,,\quad\qquad\quad \cr 
{\cal M}_{A}=\vert x\rangle\langle x\vert+\vert y\rangle\langle y\vert\,,\quad\qquad\quad 
{\cal M}_{B}=\vert p\rangle\langle p\vert+\vert q\rangle\langle q\vert\,,
\label{23}
\end{eqnarray} 
where $\vert x\rangle$, $\vert y\rangle$, $\vert p\rangle$, $\vert q\rangle$ are introduced in Eq. (\ref{3}) (note that $\vert\chi_{\alpha}\rangle\langle\chi_{\alpha}\vert =
{\cal M}{\cal A}{\cal M}$). In addition, for any $ V_{\alpha}$ such as 
\begin{eqnarray}
V_{\alpha}=c_{1}V_{(1)\alpha}+c_{2}V_{(2)\alpha}\,,\qquad
c_{1}\geq 0\,,\qquad c_{2}\geq 0\,, \qquad c_{1}+c_{2}=1\,,
\label{24}
\end{eqnarray}
inequalities (\ref{21}) and, consequently, (\ref{22}) also hold.
 
According to the definition of $V_{\alpha}$ in Eqs. (23) and (24), we have:
\begin{eqnarray}
tr\left( \rho\otimes\rho V_{\alpha}\right)= tr\left( \varrho\otimes\varrho V_{\alpha}\right)\,,\quad\cr 
\varrho = {\cal M}_{A}\otimes{\cal M}_{B}\rho{\cal M}_{A}\otimes{\cal M}_{B}\,,\quad
\label{25}
\end{eqnarray}
which means that if $V_{\alpha}$ detects the entanglement of $\rho$, it has, in fact, detected the entanglement of a two-qubit
submatrix of $\rho$. Any $\rho$ which has an entangled two-qubit submatrix is distillable~\cite{41}. So any $\rho$ which is detected by $V_{\alpha}$ is distillable.

The right hand side of the inequality (\ref{18}) is invariant under local unitary transformations~\cite{8}:
\begin{eqnarray}
tr\left( \rho\otimes\rho V_{(i)}\right)= tr\left( \rho'\otimes\rho' V_{(i)}\right)\,,\quad\cr 
\rho' = {U}_{A}\otimes{U}_{B}\rho{U}^{\dagger}_{A}\otimes{U}^{\dagger}_{B}\,,\quad\qquad
\label{25a}
\end{eqnarray}
where $U_{A}$ and $U_{B}$ are arbitrary unitary operators. This is so because $U^{\dagger}_{A}\otimes U^{\dagger}_{A} P^{A}_{\pm} U_{A}\otimes{U_{A}}=P^{A}_{\pm}$ and $U^{\dagger}_{B}\otimes U^{\dagger}_{B} P^{B}_{\pm} U_{B}\otimes{U_{B}}=P^{B}_{\pm}$. So, the choices of local bases in the definition of $V_{(i)}$ in (\ref{18}) are not important since all the choices lead to the same result. But, according to the definition of $V_{\alpha}$ in Eqs. (23) and (24), the right hand side of the inequality (\ref{22}) is not invariant under local unitary transformations. It is however expected since the $ALB_{\alpha}(\rho)$ is not invariant under such transformations either. 

Using Eqs. (23) and (24), it can be shown simply that the right hand side of the inequality (\ref{22}) is invariant under the following transformations:
\begin{eqnarray}
tr\left( \rho\otimes\rho V_{\alpha}\right)= tr\left( \rho'\otimes\rho' V_{\alpha}\right)\,,\quad\qquad\quad\cr 
\rho' = {u}_{A}\otimes{u}_{B}\rho u^{\dagger}_{A}\otimes u^{\dagger}_{B}\,, \qquad\qquad\qquad\cr
{\cal M}_{A} u_{A} {\cal M}_{A}=u_{A},\qquad\qquad u_{A}u_{A}^{\dagger}=u_{A}^{\dagger}u_{A}={\cal M}_{A}, \cr
{\cal M}_{B} u_{B} {\cal M}_{B}=u_{B}, \qquad\qquad u_{B}u_{B}^{\dagger}=u_{B}^{\dagger}u_{B}={\cal M}_{B}, \cr
\Rightarrow tr(\rho')\leq 1.\qquad\qquad\qquad\qquad
\label{25b}
\end{eqnarray} 
$\vert\chi_{\alpha}\rangle $ is also invariant, up to a phase, under the above transformations, i.e. $ u_{A} \otimes{u_{A}} \otimes u_{B} \otimes{u_{B}}\vert\chi_{\alpha}\rangle=e^{i\beta}\vert\chi_{\alpha}\rangle$ and $0\leq\beta\leq 2\pi$, but it is not so for the $ALB_{\alpha}(\rho)$. Consider the decomposition of $\rho$ into pure states as $\rho =\sum_{i}\vert\theta_{i}^{\alpha}\rangle\langle\theta_{i}^{\alpha}\vert$. From Eq. (\ref{25b}) we know that there is a decomposition of $\rho'$ into pure states as $\rho' =\sum_{i}\vert\theta_{i}^{'\alpha}\rangle\langle\theta_{i}^{'\alpha}\vert$, where $\vert\theta_{i}^{'\alpha}\rangle= u_{A} \otimes{u_{B}}\vert\theta_{i}^{\alpha}\rangle$. So, using Eq. (\ref{20}):
\begin{equation}
\sum_{i}\vert\langle\chi_{\alpha}\vert\theta_{i}^{'\alpha}\rangle\vert\theta_{i}^{'\alpha}\rangle\vert = \sum_{i}\vert\langle\chi_{\alpha}\vert\theta_{i}^{\alpha}\rangle\vert\theta_{i}^{\alpha}\rangle\vert = ALB_{\alpha}(\rho).
\label{25c}
\end{equation} 
But
\begin{equation}
\sum_{i}\vert\langle\chi_{\alpha}\vert\theta_{i}^{'\alpha}\rangle\vert\theta_{i}^{'\alpha}\rangle\vert \geq \min_{\left\lbrace \vert\psi'_{j}\rangle\right\rbrace }
\sum_{j}\vert\langle\chi_{\alpha}\vert\psi'_{j}\rangle\vert\psi'_{j}\rangle\vert = ALB_{\alpha}(\rho'),
\label{25d}
\end{equation} 
where the minimum is taken over all decompositions of $\rho'$ into pure states: $ \rho' =\sum_{j} \vert\psi'_{j}\rangle\langle\psi'_{j}\vert $. So:
\begin{equation}
 ALB_{\alpha}(\rho')\leq ALB_{\alpha}(\rho).
\label{25e}
\end{equation}
Note that expressions (\ref{22}), (\ref{25b}) and (\ref{25e}) show that $ tr(\rho\otimes \rho V_{\alpha})$ bounds the amount of $ALB^{2}_{\alpha}(\rho')$, for all possible $\rho'$ in Eq. (\ref{25b}), from below.
   
Now, using inequalities (14) and (22): 
\begin{equation}
C^{2}(\rho)\geq\sum_{\alpha}ALB_{\alpha}^{2}(\rho)\geq\sum_{\alpha}tr\left( \rho\otimes\rho V_{\alpha}\right)\,,
\label{26}
\end{equation}
where the summation is only over those $\alpha$ for which $ tr\left( \rho\otimes\rho V_{\alpha}\right)\geq 0$.

\textit{Example 1.} In a $d\times d$ dimensional Hilbert space, isotropic states are defined 
as~\cite{2}:
\begin{eqnarray}
\rho_{_{F}}= \frac{1-F}{d^{2}-1}\left(  I - \vert\phi^{+}\rangle\langle\phi^{+}\vert \right) + F\vert\phi^{+}\rangle\langle\phi^{+}\vert \,,\cr 
\vert\phi^{+}\rangle =\sum_{i=1}^{d}\frac{1}{\sqrt{d}}\vert i_{A}i_{B}\rangle\, ,\qquad\qquad\cr 
0\leq F\leq1\,, \qquad F=\langle\phi^{+}\vert\rho_{_{F}}\vert\phi^{+}\rangle\,.\quad
\label{27}
\end{eqnarray}
The concurrence of $\rho_{_{F}}$ is known and we have ~\cite{33}:
\begin{equation}
C\left( \rho_{_{F}}\right)= max \left\lbrace 0  ,  \sqrt{\frac{2d}{d-1}}\left(F-\frac{1}{d}\right)\right\rbrace \,.
\label{28} 
\end{equation}
If we rewrite $\rho_{_{F}}$ as
\begin{eqnarray*}
\rho_{_{F}}=\frac{1-F}{d^{2}-1}I+\frac{Fd^{2}-1}{d^{2}-1}\vert\phi^{+}\rangle\langle\phi^{+}\vert 
\equiv gI+h\vert\phi^{+}\rangle\langle\phi^{+}\vert\,,\qquad\qquad\quad
\end{eqnarray*}
then:
 \begin{eqnarray}
 tr\left( \rho_{_{F}}\otimes\rho_{_{F}} V_{(i)}\right)
 =2d\left( d-1\right) \left[ \frac{h^{2}}{d^{2}} -dg^{2}- \frac{2}{d}gh\right]\,.
 \label{29}
 \end{eqnarray}
 In Eq. (\ref{23}), if we choose $ \left\lbrace x=p, y=q\right\rbrace  $, then:
 \begin{eqnarray*}
tr\left( \rho_{_{F}}\otimes\rho_{_{F}} V_{\alpha}\right) = 4\left[ \frac{h^{2}}{d^{2}} -2g^{2}- \frac{2}{d}gh\right]\,,
\end{eqnarray*}
and the expectation values of other $ V_{\alpha} $ are not positive. Since the case $\left\lbrace x=p, y=q\right\rbrace $ occurs $n=\frac{d(d-1)}{2}$ times in a $ d\times d $ dimensional system, we have:
\begin{eqnarray}
tr\left(\rho_{_{F}}\otimes\rho_{_{F}}\sum_{\alpha}V_{\alpha}\right)=2d(d-1)\left[ \frac{h^{2}}{d^{2}}-2g^{2}-\frac{2}{d}gh\right]\,,
 \label{30} 
\end{eqnarray} 
where the summation is only over those $V_{\alpha}$ for which $\left\lbrace x=p, y=q\right\rbrace $. For $d>2$, Eq. (\ref{30})
gives a better result than Eq. (\ref{29}) (Fig. 1). For $d=2$ both give the same result, as we expect from Eq. (\ref{23}).

\begin{figure} [htbp]
\vspace*{13pt}
\centerline{\psfig{file=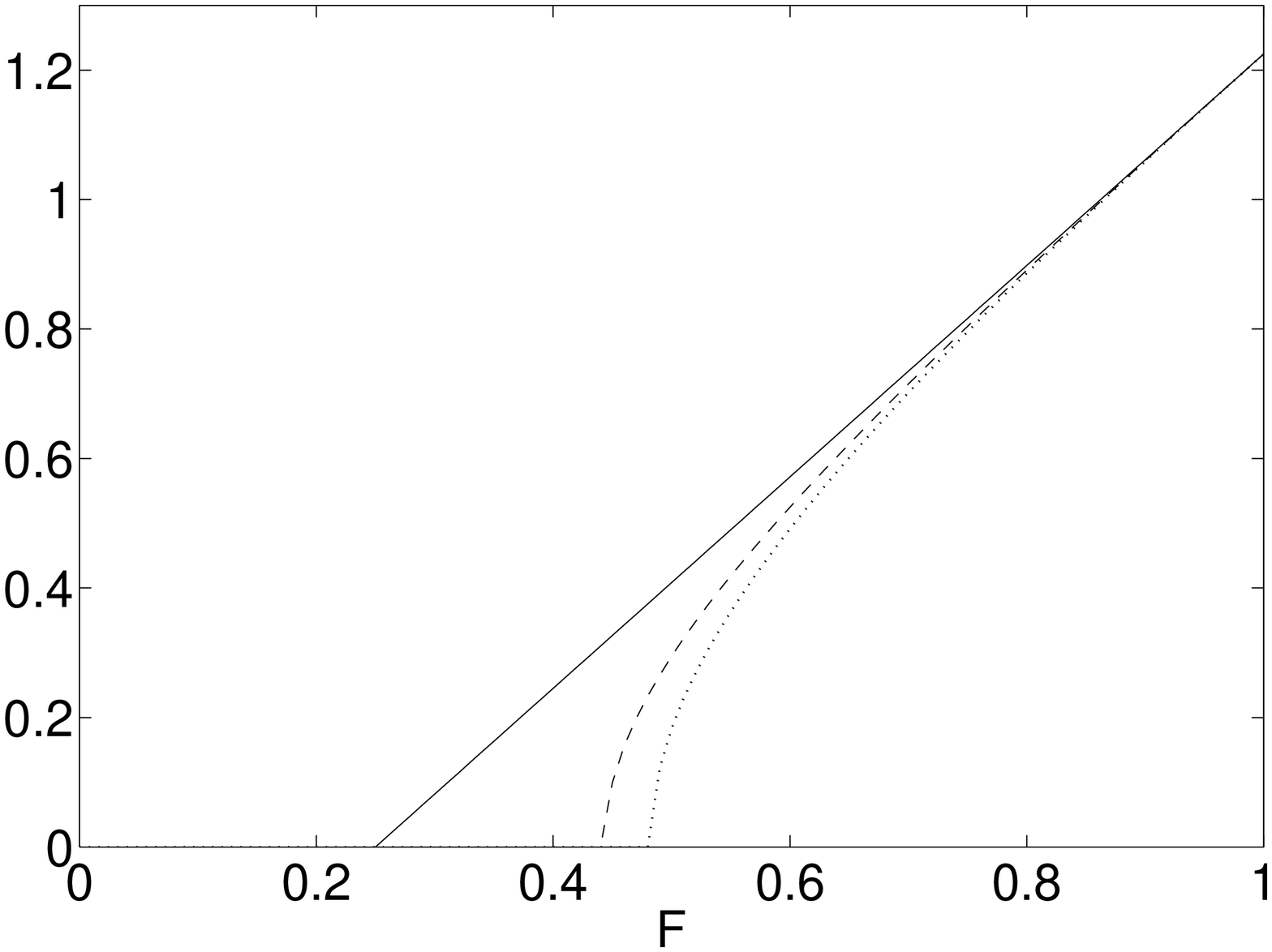, width=8.2cm}} 
\vspace*{13pt}
\fcaption{\label{fig1}
Comparison of Eqs. (\ref{29}), dotted line, and (\ref{30}), dashed line, for $ d=4 $. The solid 
line is the exact value of concurrence, Eq. (\ref{28}). The lower bounds given by $ V_{(i)} $ and $ \sum_{\alpha}V_{\alpha} $ are set to zero when the right hand sides of Eqs. (\ref{29}) and (\ref{30})
are less than zero.}
\end{figure}


\section{Measurable Lower Bounds in terms of One Copy of $\rho$}
\noindent
From the experimental point of view, any lower bound which is defined in terms of the expectation value of an observable with respect to two identical copies of $\rho$, encounters, at least, two problems. First, for measuring $V_{(i)}$ or $V_{\alpha}$
we need to measure in an entangled basis in both parts A and B. Measuring in an entangled basis is more difficult than measuring in a separable one ~\cite{12}. Second, it is not clear that the state which enters the measuring devices is really as $\rho\otimes\rho$ even if we can produce such state at the source place 
~\cite{42,10}. So, having lower bounds in terms of the expectation value of an observable with respect to \textit{one} copy of $\rho$ is more desirable.

Using:
\begin{eqnarray}
C(\rho)C(\sigma)\geq tr\left( \rho\otimes\sigma V_{(i)}\right)\,, \qquad i=1, 2\,;\cr 
\Rightarrow C(\rho)\geq\frac{1}{C(\sigma)}tr\left(\rho\otimes\sigma V_{(i)}\right)\,,\qquad
\label{31} 
\end{eqnarray}
for arbitrary $\rho$ and $\sigma$, F. Mintert has introduced the following measurable lower bound on $C(\rho)$~\cite{14}:
\begin{eqnarray}
C(\rho)\geq -tr\left( \rho W_{\sigma}\right)\,,\qquad 
W_{\sigma}=\frac{-1}{C(\sigma)}tr_{2}\left( I\otimes\sigma V_{(i)}\right)\,,
\label{32}  
\end{eqnarray}
where $\sigma$ is a pre-determined entangled state and the partial trace is taken over the second copy of 
$ \mathcal{H}_{A}\otimes \mathcal{H}_{B}\ $. If $C(\sigma)$ is not computable, which is the
case for almost all
mixed $ \sigma $, an upper bound of $C(\sigma)$ can be used in the 
definition of  $W_{\sigma}$. From inequality (\ref{32}), it is obvious that for any separable state: $tr\left(\rho_{s}W_{\sigma}\right)\geq 0$. If, at least, for one entangled state  
$tr\left(\rho_{e}W_{\sigma}\right)< 0$, then $W_{\sigma}$ is an entanglement witness~\cite{2}.

We can, also, construct measurable lower bounds in terms of one copy of $\rho$ by using inequality (\ref{21}).
Suppose that the decomposition of $\sigma$ which gives the minimum in Eq. (\ref{9}) is 
$\sigma =\sum_{j}\vert\gamma_{j}^{\alpha}\rangle\langle\gamma_{j}^{\alpha}\vert$, i.e.:
\begin{equation}
ALB_{\alpha}(\sigma)=\sum_{j}\vert\langle\chi_{\alpha}\vert\gamma_{j}^{\alpha}\rangle\vert\gamma_{j}^{\alpha}\rangle\vert\,.
\label{33}
\end{equation}
Using expressions (\ref{20}), (21) and (\ref{33}):
\begin{eqnarray*}
\left[ ALB_{\alpha}(\rho)\right]\left[ ALB_{\alpha}(\sigma)\right]
=\sum_{ij}\vert\langle\chi_{\alpha}\vert\theta_{i}^{\alpha}\rangle\vert\theta_{i}^{\alpha}\rangle\vert
\vert\langle\chi_{\alpha}\vert\gamma_{j}^{\alpha}\rangle\vert\gamma_{j}^{\alpha}\rangle\vert
\cr 
\geq \sum_{ij}\langle\theta_{i}^{\alpha}\vert\langle\gamma_{j}^{\alpha}\vert V_{\alpha}\vert\theta_{i}^{\alpha}\rangle\vert\gamma_{j}^{\alpha}\rangle =-tr\left( \rho W_{\sigma\alpha}^{\prime}\right)\,,\quad\cr 
 W_{\sigma\alpha}^{\prime}= -tr_{2}\left(I\otimes\sigma V_{\alpha}\right)\,.\qquad\qquad\qquad   
\end{eqnarray*}
So:
\begin{eqnarray}
ALB_{\alpha}(\rho)\geq -tr\left( \rho W_{\sigma\alpha}\right)\,, \qquad\qquad
W_{\sigma\alpha}=\frac{1}{ALB_{\alpha}(\sigma)}W_{\sigma\alpha}^{\prime}\,,
\label{34}
\end{eqnarray}
where $\sigma$ is a pre-determined entangled state for which $ALB_{\alpha}(\sigma)>0$. Note that, in contrast to
$C(\sigma)$, $ALB_{\alpha}(\sigma)$ is always computable, so we never need to use an upper bound of it in the definition of $W_{\sigma\alpha}$.
In addition, it can be shown simply that 
\begin{equation}
tr\left( \rho W_{\sigma\alpha}\right) = tr\left( \varrho W_{\sigma\alpha}\right)\,,
\label{35}
\end{equation}
where $\varrho$ is defined in Eq. (\ref{25}). So any $\rho$ which is detected by $W_{\sigma\alpha}$ is distillable. Also, using inequalities (\ref{14}) and (\ref{34}):
\begin{equation}
C^{2}(\rho)\geq\sum_{\alpha}\left[ALB_{\alpha}(\rho)\right]^{2}\geq\sum_{\alpha}\left[tr\left(\rho W_{\sigma\alpha}\right) \right] ^{2}\,, 
\label{36}
\end{equation}
where the summation is over those $\alpha$ for which $ tr\left(\rho W_{\sigma\alpha}\right)\leq 0$.

For isotropic states,using expressions (\ref{32}) or (\ref{36}) (by choosing $\sigma =\vert\phi^{+}\rangle\langle\phi^{+}\vert$) gives the exact value of $C(\rho_{_{F}})$ for arbitrary $d$. In the following, we give an example for which the expression (\ref{36}) gives better results than the expression (\ref{32}).

\textit{Example 2.} Consider a two-qutrit system which is initially in the pure state
\begin{equation}
\vert\Phi\rangle =\sqrt{\lambda_{0}}\vert 01\rangle +\sqrt{\lambda_{1}}\vert 12\rangle+\sqrt{\lambda_{2}}\vert 20\rangle\,,
\label{37}
\end{equation}
and its time evolution is given by the following Master equation~\cite{14}:
\begin{eqnarray}
\dot\rho={\cal L}\rho\,,\qquad\qquad\cr 
{\cal L}={\cal L}_{A}\otimes 1_{B}+1_{A}\otimes {\cal L}_{B}\,,
\label{38}
\end{eqnarray} 
where ${\cal L}_{A/B}$, for a one-qutrit $\rho_{A/B}$, is 
\begin{eqnarray*}
{\cal L}_{A/B}=\frac{\Gamma}{2}\left(2\gamma\rho_{A/B}\gamma^{\dagger} -\rho_{A/B}\gamma^{\dagger}\gamma -
\gamma^{\dagger}\gamma\rho_{A/B}\right)\,, 
\end{eqnarray*}
and $\gamma$ is the coupling matrix for the spontaneous decay:
\begin{eqnarray*}
\gamma=\left(
\begin{array}{ccc}
0 & 0 & 0  \\
\sqrt{2} & 0 & 0 \\
0 & 1 & 0 
\end{array}
\right)\, .  \label{reduced}
\end{eqnarray*}
To construct $W_{\sigma\alpha}$ in expression (\ref{34}) and $W_{\sigma}$ in expression (\ref{32}), we choose
\begin{eqnarray}
\sigma =\vert\Phi_{ME}\rangle\langle\Phi_{ME}\vert\,,\qquad\qquad\cr 
\vert\Phi_{ME}\rangle =\frac{1}{\sqrt{3}}\left(\vert 01\rangle +\vert 12\rangle +\vert 20\rangle\right)\,.
\label{39} 
\end{eqnarray} 
It can be shown simply that for three $\vert\chi_{\alpha}\rangle$, for which $\left\lbrace p=x\oplus 1, q=y\oplus 1\right\rbrace $
($\oplus$ is the sum modulo 3), $ALB_{\alpha}(\sigma)=2/3 $, and 
$ALB_{\alpha}(\sigma)=0$ for other $\vert\chi_{\alpha}\rangle$. So, using expression (\ref{34}), we can construct three $W_{\sigma\alpha}$ as ($x=0,1,2$ and $y=x\oplus 1$):
\begin{eqnarray}
W_{\sigma\alpha}=\vert x,y\oplus 1\rangle\langle x,y\oplus 1\vert +\vert y,x\oplus 1\rangle\langle y,x\oplus 1\vert -\vert x,x\oplus 1\rangle\langle y,y\oplus 1\vert-\vert y,y\oplus 1\rangle\langle x,x\oplus 1\vert\cr =\vert x,y\oplus 1\rangle\langle x,y\oplus 1\vert +\vert y,x\oplus 1\rangle\langle y,x\oplus 1\vert -\frac{1}{2}\left(\sigma_{1}^{xy}\otimes\sigma_{1}^{x\oplus 1,y\oplus 1}-\sigma_{2}^{xy}\otimes\sigma_{2}^{x\oplus 1,y\oplus 1}\right)\,,\cr 
\sigma_{1}^{ab}=\vert a\rangle\langle b\vert +\vert b\rangle\langle a\vert\,,\qquad\quad\qquad 
\sigma_{2}^{ab}=-i\left( \vert a\rangle\langle b\vert -\vert b\rangle\langle a\vert\right)\,.\qquad\qquad
\label{40} 
\end{eqnarray}
Also, using expression (\ref{32}), we can show that:
\begin{equation}
W_{\sigma}=\frac{1}{\sqrt{3}}\sum_{\alpha =1}^{3}W_{\sigma\alpha}\,.
\label{41}
\end{equation}  
As we can see from Eqs. (\ref{40}) and (\ref{41}), the number of local observables needed for measuring $W_{\sigma}$ or three 
$W_{\sigma\alpha}$ is the same and is equal to 12, which is less than what
is needed for a full tomography. Also, note that $ \left\lbrace \vert l,m\oplus 1\rangle\right\rbrace $ is an 
orthonormal basis of $\mathcal{H}_{A}\otimes \mathcal{H}_{B} $. So, at least from the theoretical
point of view, all the observables 
$\vert l,m\oplus 1\rangle\langle l,m\oplus 1\vert$ can be measured using only one set up. In such cases, for measuring $W_{\sigma}$
or three $W_{\sigma\alpha}$, we only need 7 different set up of local measurements. The comparison of the results of 
inequalities (\ref{32}) and (\ref{36}), for two typical $\left\lbrace \lambda_{i}\right\rbrace $, is given in Fig. 2.

\begin{figure} [htbp]
\vspace*{13pt}
\centerline{\psfig{file=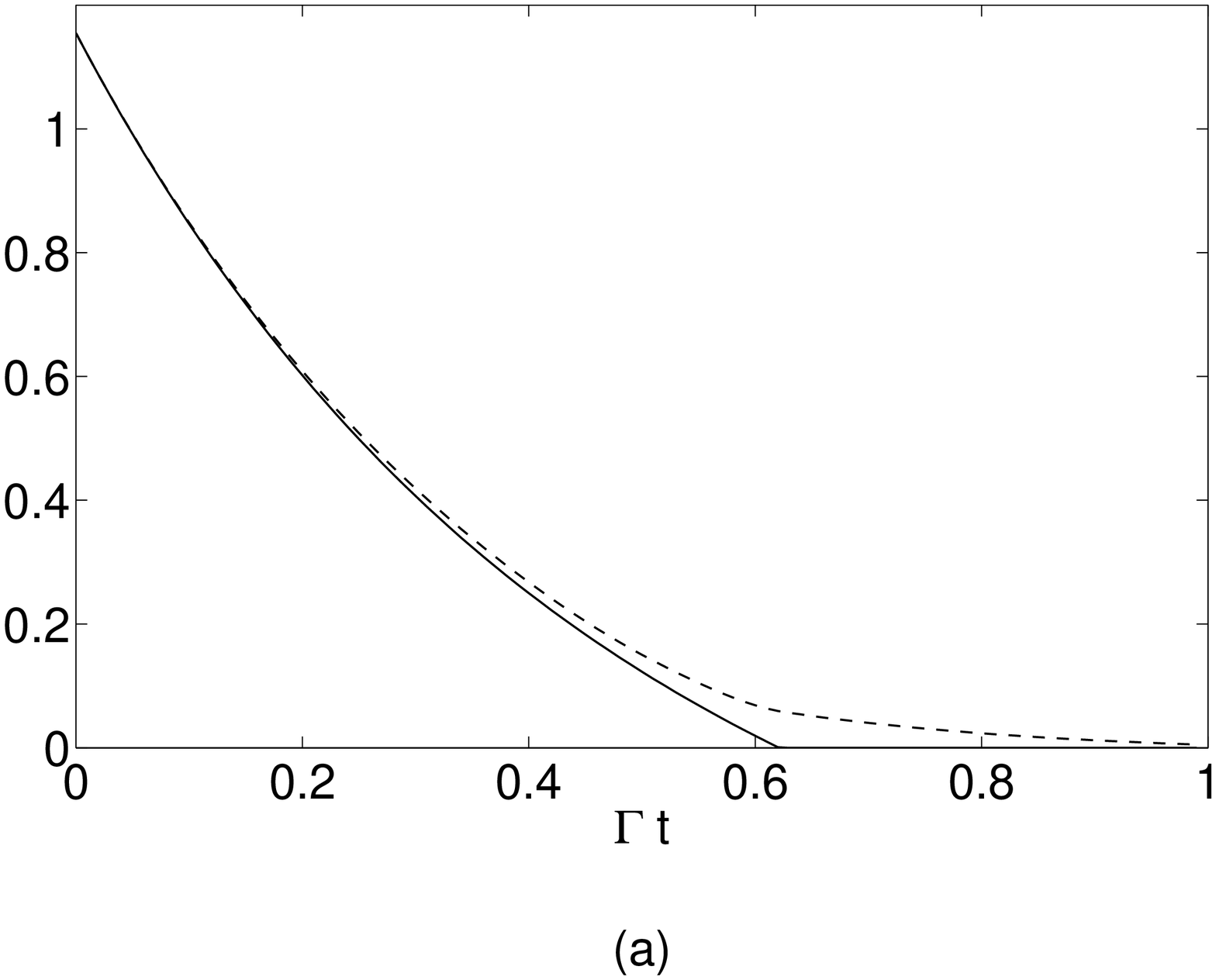, width=8.2cm}} 
\centerline{\psfig{file=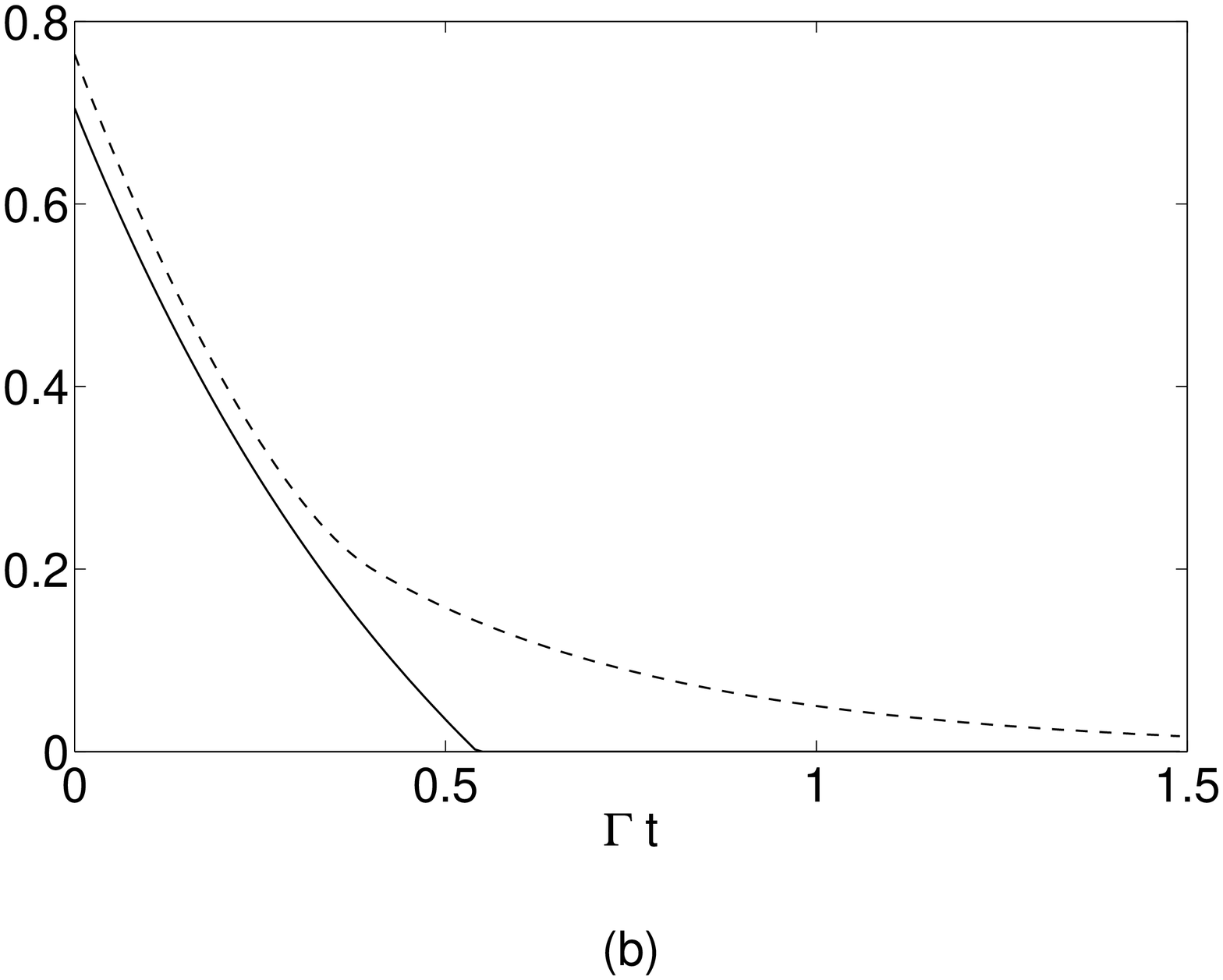, width=8.2cm}} 
\vspace*{13pt}
\fcaption{\label{fig2}
Comparing the lower bounds given by (\ref{32}), solid line, and (\ref{36}), dashed line, for two typical $\left\lbrace \lambda_{i}\right\rbrace $: a) 
$ \lambda_{i}=1/3 $; b) $ \left\lbrace \lambda_{0}=1/12,\lambda_{1}=5/6,\lambda_{2}=1/12\right\rbrace $. When
the lower bound given by $ W_{\sigma} $ is less than zero, we set it to zero.}
\end{figure}


\section{Extending to Multipartite Systems}
\noindent
In a bipartite system, any Hermitian operator which, for arbitrary $\vert\psi\rangle$ and $\vert\varphi\rangle$, 
satisfies the inequality
\begin{equation}
C(\psi)C(\varphi)\geq\langle\psi\vert\langle\varphi\vert V\vert\psi\rangle\vert\varphi\rangle\,, 
\label{41a}
\end{equation}
gives a measurable lower bound on $C^{2}(\rho)$, i.e. 
$C^{2}(\rho)\geq tr\left( \rho\otimes\rho V\right)$ ~\cite{10}. This can be proved simply by writing $\rho$ in terms of its extremal decomposition $\rho =\sum_{j}\vert\xi_{j}\rangle\langle\xi_{j}\vert$. In~\cite{17} it was shown 
how to use such $V$ to construct measurable lower bounds for multipartite concurrence. Following a similar procedure, we construct measurable lower bounds on multipartite concurrence using $V_{\alpha}$. As the previous sections, we will use the inequality (\ref{21}) instead of the inequality (\ref{41a}). In other words, we will work with the algebraic lower bounds of $C(\rho)$
rather than the concurrence itself.

The concurrence of an N-partite pure state $\vert\Psi\rangle$, $\vert\Psi\rangle\in{\mathcal H}_{A_{1}}\otimes \cdots\otimes{\mathcal H}_{A_{N}}$, is defined as~\cite{30}:
\begin{equation}
C(\Psi)=2^{1-\frac{N}{2}}\sqrt{\sum_{l}C_{l}^{2}(\Psi)}\,,
\label{42}
\end{equation}
where $\sum_{l}$ is the summation over all possible subdivisions of ${\mathcal H}_{A_{1}}\otimes \cdots\otimes{\mathcal H}_{A_{N}}  $  into two subsystems, and $C_{l}$ is the related bipartite concurrence. For example, for a 3-partite system we have three  $C_{l}$, namely $C_{1,23},C_{12,3}$ and $C_{13,2}$. As 
before we have:
\begin{eqnarray}
C_{l}^{2}\left( \Psi\right) =\langle\Psi\vert\langle\Psi\vert{\cal A}_{l}\vert\Psi\rangle\vert\Psi\rangle\,,\qquad
{\cal A}_{l}=\sum_{\alpha_{l}}\vert\chi_{\alpha_{l}}\rangle\langle\chi_{\alpha_{l}}\vert\,,
\label{43}
\end{eqnarray}
where $\vert\chi_{\alpha_{l}}\rangle$ are the same as $\vert\chi_{\alpha}\rangle$ which have been defined in Eq. (\ref{3}).
Obviously, they are constructed according to the related subdivision denoted by $l $. So: 
\begin{eqnarray}
C(\Psi)=2^{1-\frac{N}{2}}\sqrt{\sum_{l,\alpha_{l}}\vert\langle\chi_{\alpha_{l}}\vert\Psi\rangle\vert\Psi\rangle\vert^{2}} 
=2^{1-\frac{N}{2}}\sqrt{\sum_{\gamma}\vert\langle\chi_{\gamma}\vert\Psi\rangle\vert\Psi\rangle\vert^{2}}\,,
\label{44}
\end{eqnarray}
where instead of $l$ and $\alpha_{l}$ we have used a collective index $\gamma$. From now on, everything is as the bipartite case, except that we deal with the summation over $\gamma$ instead of $\alpha$. The definition
 of concurrence for mixed states is as follows:
\begin{eqnarray}
C(\rho)=\min_{\left\lbrace \vert\psi_{i}\rangle\right\rbrace} \sum_{i}C(\psi_{i}) 
=\min_{\left\lbrace \vert\psi_{i}\rangle\right\rbrace} \sum_{i}2^{1-\frac{N}{2}}
\sqrt{\langle\psi_{i}\vert\langle\psi_{i}\vert{\cal A}^{\prime}\vert\psi_{i}\rangle\vert\psi_{i}\rangle}\,,\cr 
{\cal A}^{\prime}=\sum_{\gamma}\vert\chi_{\gamma}\rangle\langle\chi_{\gamma}\vert\,,\qquad\qquad\qquad\qquad
\label{45}
\end{eqnarray}
where the minimization is over all decompositions of $\rho$ into subnormalized states $\vert\psi_{i}\rangle$:
$\rho =\sum_{i}\vert\psi_{i}\rangle\langle\psi_{i}\vert$. It is worth noting that $C(\rho)$, as difined in Eq. (\ref{45}), 
is an entanglement monotone for the multipartite case too~\cite{42b}.

If we define $ \vert\chi_{\upsilon}^{\prime}\rangle = \sum_{\gamma}U^{\prime}_{\upsilon\gamma}
\vert\chi_{\gamma} \rangle$, where $ U ^{\prime}$ is a unitary matrix, then $ {\cal A}^{\prime}=\sum_{\gamma}\vert\chi_{\gamma}\rangle\langle\chi_{\gamma}\vert = \sum_{\gamma}\vert\chi^{\prime}_{\gamma}\rangle\langle\chi^{\prime}_{\gamma}\vert $. So, by similar reasoning leading to inequality (\ref{13}), 
we have:
\begin{eqnarray}
C(\rho)\geq LB_{\tau}(\rho)=\min_{\left\lbrace \vert\psi_{i}\rangle\right\rbrace} \sum_{i}2^{1-\frac{N}{2}}\vert\langle\tau\vert\psi_{i}\rangle\vert\psi_{i}\rangle\vert\,,\cr 
\vert\tau\rangle\equiv\vert\chi^{\prime}_{1}\rangle =\sum_{\gamma}z_{\gamma}^{\ast}\vert\chi_{\gamma}\rangle\, ,\qquad\sum_{\gamma}\vert z_{\gamma}\vert^{2}=1\,.\qquad
\label{46}
\end{eqnarray}
As before, in contrast to $C(\rho)$, $LB_{\tau}(\rho)$ is always computable. We also have:
\begin{eqnarray}
C^{2}(\rho)\geq\sum_{\gamma}\left[LB_{\gamma}(\rho)\right]^{2}\,,\qquad 
LB_{\gamma}(\rho)=\min_{\left\lbrace \vert\psi_{i}\rangle\right\rbrace} 2^{1-\frac{N}{2}}\sum_{\gamma}\vert\langle\chi_{\gamma}^{\prime}\vert\psi_{i}\rangle\vert\psi_{i}\rangle\vert\,.
\label{47}
\end{eqnarray} 
The above expression is the counterpart of the inequality (\ref{14}) for the multipartite case. What was proved in~\cite{39}, neglecting an unimportant constant in the definition of $C(\rho)$, is, in fact, the inequality (\ref{47}) for the special case of   ${\vert\chi_{\gamma}^{\prime}\rangle}={\vert\chi_{\gamma}\rangle}$ 
(see Eqs. (16) and (17)).

According to the inequality (\ref{21}),for any $\vert\chi_{\gamma}\rangle$:
\begin{equation}
\vert\langle\chi_{\gamma}\vert\psi\rangle\vert\psi\rangle\vert\vert\langle\chi_{\gamma}
\vert\varphi\rangle\vert\varphi\rangle\vert\geq\langle\psi\vert\langle\varphi\vert V_{\gamma}\vert\psi\rangle\vert\varphi\rangle\,,
\label{48}
\end{equation}
where $ V_{\gamma} $ are the same as $ V_{\alpha} $ introduced in Eqs. (23) and (24), defined 
according to the related $ \vert\chi_{\gamma}\rangle $. So:
\begin{equation}
C^{2}(\rho)\geq 2^{2-N}\sum_{\gamma}tr\left(\rho\otimes\rho V_{\gamma}\right)\,,
\label{49} 
\end{equation}
where the summation is over those $\gamma$ for which $tr\left(\rho\otimes\rho V_{\gamma}\right)\geq 0$.
Also, we have:
\begin{eqnarray}
C^{2}(\rho)\geq\sum_{\gamma}\left[ tr\left(\rho W_{\sigma\gamma}\right)\right]^{2}\,,\qquad 
W_{\sigma\gamma}=\frac{-2^{2-N}}{ALB_{\gamma}(\sigma)}tr_{2}\left(I\otimes\sigma V_{\gamma}\right)\,,
\label{50}  
\end{eqnarray}
where $\sigma$ is a pre-determined density operator with $ALB_{\gamma}(\sigma)>0$, and the summation is over those $\gamma$ for which $tr\left(\rho W_{\sigma\gamma}\right)\leq 0$.


\section{Summery and Discussion}
\noindent
Inequality (\ref{21}) is the main relation of this paper. Using this expression, we have constructed measurable lower bounds on concurrence in term of
both one copy or two identical copies of $\rho$. We have proved that the inequality (\ref{21}) holds for $V_{\alpha}$ introduced in Eq. (\ref{23}). 
Now verifying  whether it is possible to find $V^{\prime}_{\alpha}$ for which (\ref{21}) holds for arbitrary $\vert\chi_{\alpha}^{\prime}\rangle$ is
valuable.

Our measurable bounds are related to the $ALB_{\alpha}(\rho)$ rather than the concurrence itself, as we have seen in expressions (\ref{22}) and (\ref{34}).
So we can use (\ref{14}) to get the relations (\ref{26}) and (\ref{36}). Inequality (\ref{26})(Inequality (\ref{36})) has this advantage that we can omit the summation
over those $\alpha$ for which $ tr\left( \rho\otimes\rho V_{\alpha}\right)\leq 0$ ($ tr\left(\rho W_{\sigma\alpha}\right)\geq 0$). This useful 
property can help us to achieve better results in detecting the entanglement. As an example, $W_{\sigma}$ in Eq. (\ref{41}) is, up to a constant, the
summation of three $W_{\sigma\alpha}$. Now, using expression (\ref{36}), we can omit each $W_{\sigma\alpha}$ for which
$ tr\left(\rho W_{\sigma\alpha}\right)\geq 0$; But, using $W_{\sigma}$, we can not omit any $W_{\sigma\alpha}$ in Eq. (\ref{41}). So, as it is shown
 in Fig. 2, the ability of $W_{\sigma}$ in detecting the entanglement reduces more rapidly than the three distinct $W_{\sigma\alpha}$. 
 
Bounds obtained from $V_{\alpha}$ or $W_{\sigma\alpha}$ are always less than or equal to the $ALB_{\alpha}(\rho)$. In addition, we have shown that these bounds can not detect bound entangled states. So $ALB_{\alpha}(\rho)>0$ and $N(\rho)>0$ ($N(\rho)$ is the negativity of the system ~\cite{43}) are two necessary conditions for detection of the entanglement by $V_{\alpha}$ or $W_{\sigma\alpha}$.
However, the ability of these bounds and also comparing them with other observable bounds, especially those introduced in ~\cite{8,14}, need further studies. For example, in the definition of $W_{\sigma\alpha}$, mixed states $\sigma$ can be used simply instead of pure states $\sigma$ 
since $ALB_{\alpha}(\sigma)$ is always computable. Studying the above case seems interesting.

At last, in section V, we have generalized our measurable bounds to the multipartite case. The applicability of these bounds also needs further studies.


\nonumsection{Acknowledgements}
\noindent
We would like to thank the anonymous referees for their helpful suggestions and improvements.

\nonumsection{References}

\appendix

\noindent
In this appendix, we prove inequality (21) for $V_{\alpha}$ introduced in Eq. (\ref{23}). We prove it for $V_{(2)\alpha}$; the case of $V_{(1)\alpha}$ can be done analogously.

Any arbitrary $\vert\psi\rangle$ and $\vert\varphi\rangle$ can be decomposed in a separable basis of $ \mathcal{H}_{A}\otimes \mathcal{H}_{B}\ $, like ${\vert i_{A}\rangle\vert j_{B}\rangle}$, as: 
\begin{eqnarray*}
\vert\psi\rangle =\sum_{ij}\psi_{ij}\vert i_{A}j_{B}\rangle\,,\cr 
\vert\varphi\rangle =\sum_{ij}\varphi_{ij}\vert i_{A}j_{B}\rangle\,.
\end{eqnarray*}
Now, from Eq. (\ref{23}), we have: 
\begin{eqnarray}
\langle\psi\vert\langle\varphi\vert V_{(2)\alpha}\vert\psi\rangle\vert\varphi\rangle =\qquad\qquad\qquad\qquad\cr 
2\left[-\vert\psi_{xq}\varphi_{yq}-\psi_{yq}\varphi_{xq}\vert^{2}-\vert\psi_{xp}\varphi
_{yp}-\psi_{yp}\varphi_{xp}\vert^{2}+AA\right]\,,\quad\cr 
AA= 
-2 Re\left(\psi_{xp}\varphi_{yq}\psi_{xq}^{\ast}\varphi_{yp}^{\ast}\right)
-2 Re\left(\psi_{yp}\varphi_{xq}\psi_{yq}^{\ast}\varphi_{xp}^{\ast}\right)\qquad\cr 
+2 Re\left(\psi_{xp}\varphi_{yq}\psi_{yq}^{\ast}\varphi_{xp}^{\ast}\right)+
2 Re\left(\psi_{xq}\varphi_{yp}\psi_{yp}^{\ast}\varphi_{xq}^{\ast}\right)\,.\qquad
\label{a1} 
\end{eqnarray}
Also for $\vert\chi_{\alpha}\rangle =\left(\vert xy\rangle -\vert yx\rangle\right)_{A}\left(\vert pq\rangle -\vert qp\rangle\right)_{B}$ we have:
\begin{eqnarray}
\vert\langle\chi_{\alpha}\vert\psi\rangle\vert\psi\rangle\vert\vert\langle\chi_{\alpha}
\vert\varphi\rangle\vert\varphi\rangle\vert\qquad\qquad\cr
=4\vert\left(\psi_{xp}\psi_{yq}-\psi_{xq}\psi_{yp}\right)
\left(\varphi_{xp}\varphi_{yq}-\varphi_{xq}\varphi_{yp}\right)\vert\cr 
\equiv 4\vert BB\vert\,.\qquad\qquad\qquad\quad
\label{a2}
\end{eqnarray}
To get the inequality (21), we must show: 
\begin{eqnarray}
AA\leq 2\vert BB\vert +\vert\psi_{xq}\varphi_{yq}-\psi_{yq}\varphi_{xq}\vert^{2}\cr 
+\vert\psi_{xp}\varphi_{yp}-\psi_{yp}\varphi_{xp}\vert^{2}\,.
\label{a3}
\end{eqnarray}
If we have: 
\begin{eqnarray}
AA\leq 2\vert BB\vert +2\vert CC\vert\,,\qquad\quad\cr 
 CC=\left(\psi_{xq}\varphi_{yq}-\psi_{yq}\varphi_{xq}\right)\left( \psi_{xp}\varphi_{yp}-\psi_{yp}\varphi_{xp}\right)\,,
\label{a4} 
\end{eqnarray}
then inequality (A.3) holds. To get the inequality (\ref{a4}), it is sufficient to have:
\begin{eqnarray}
\frac{AA}{2}\leq \vert BB +CC\vert\quad\qquad\quad\cr  
=\vert \left(\psi_{xp}\varphi_{yq}-\psi_{yp}\varphi_{xq}\right)\left( \psi_{yq}^{\ast}\varphi_{xp}^{\ast}-\psi_{xq}^{\ast}\varphi_{yp}^{\ast}\right)\vert\,.
\label{a5}
\end{eqnarray}
But, the above expression holds since for any complex number $z$, we have $Re(z)\leq\vert z\vert$, which completes the proof.


\noindent

\begin{thebibliography}{000}
\bibitem{1}
O. Guhne and G. Toth (2009), {\it Entanglement detection}, 
Phys. Rep., 474, pp. 1-75.

\bibitem{2}
R. Horodecki \textit{et al.} (2009), {\it Quantum entanglement}, Rev. Mod. Phys., 81, pp. 865-942.

\bibitem{3}
M. A. Nielsen and I. L. Chuang (2000), {\it Quantum Computation and Quantum Information},
Cambridge University Press (Cambridge).

\bibitem{4} O. Guhne \textit{et al.} (2007), {\it Estimating entanglement measures in experiments}, Phys. Rev. Lett.
, 98, 110502.

\bibitem{5} O. Guhne \textit{et al.} (2008), {\it Lower bounds on  entanglement measures from incomplete information}, Phys. Rev. A, 77, 052317.

\bibitem{6} J. Eisert \textit{et al.} (2007), {\it Quantitative entanglement witnesses}, New. J. Phys., 9, 46.

\bibitem{7} K. M. R. Audenaert and M. B. Plenio (2006), {\it When are correlations quantum?-verification and quantification of entanglement by simple measurements}, New. J. Phys., 8, 266.

\bibitem{8} F. Mintert and A. Buchleitner (2007), {\it Observable entanglement measures for mixed quantum states}, Phys. Rev.
Lett., 98, 140505.

\bibitem{9} C. Schmid \textit{et al.} (2008), {\it Experimental direct observation of mixed state entanglement}, Phys. Rev. Lett., 101, 260505.

\bibitem{10} F. Mintert (2007), {\it Entanglement measures as physical observables}, Appl. Phys. B, 89, pp. 493-497.

\bibitem{11} S. P. Walborn \textit{et al.} (2006), {\it Experimental determination  of entanglement with a single measurement}, Nature (London), 440, pp. 1022-1024.

\bibitem{12} S. P. Walborn \textit{et al.} (2007), {\it Experimental determination  of entanglement by a projective measurement}, Phys. Rev. A, 75, 032338.

\bibitem{13} Z. Ma \textit{et al.} (2009), {\it Bounds of concurrence and their relation with fidelity and frontier states}, Phys. Lett. A, 373,
pp. 1616-1620.

\bibitem{14} F. Mintert (2007), {\it Concurrence via entanglement witnesses}, Phys. Rev. A, 75, 052302.

\bibitem{15} C.-J. Zhang \textit{et al.} (2008), {\it Observable estimation of entanglement for arbitrary finite-dimensional mixed states}, Phys. Rev. A, 78,
042308.

\bibitem{15a} Y.-F. Huang \textit{et al.} (2009), {\it Experimental measurement of lower and upper bounds of concurrence for mixed quantum quantum states}, Phys. Rev. A, 79, 052338. 

\bibitem{16} L. Aolita and F. Mintert (2006), {\it Measuring multipartite concurrence with a single factorizable observable}, Phys. Rev. Lett., 97, 050501.

\bibitem{17} L. Aolita \textit{et al.} (2008), {\it Scalable method to estimate experimentally the entanglement of multipartite systems}, Phys. Rev. A, 78,
022308.

\bibitem{19} H.-P. Breuer (2006), {\it Separability criteria and bounds for entanglement measures}, J. Phys. A: Math. Gen., 39, pp. 11847-11860.

\bibitem{21} C.-S. Yu \textit{et al.} (2008), {\it Measurable concurrence of mixed states}, Phys. Rev. A, 77, 012305.

\bibitem{27} J. I. de Vicente (2008), {\it Lower bounds on concurrence and separability conditions}, Phys. Rev. A, 75, 052320; J. I. de Vicente (2008), {\it Erratum: Lower bounds on concurrence and separability conditions}, Phys. Rev. A, 77, 039903(E).

\bibitem{28} C.-J. Zhang \textit{et al.} (2007), {\it Optimal entanglement witnesses based on local orthogonal observables}, Phys. Rev. A, 76, 012334.

\bibitem{28p} M. Li \textit{et al.} (2008), {\it Separability and entanglement of quantum states based on covariance matrices}, J. Phys. A: Math. Theor., 41, 202002.

\bibitem{29} R. Augusiak and M. Lewenstein (2009), {\it Towards measurable bounds on entanglement measures}, Quantum Inf. Process., 8, pp. 493-521.

\bibitem{29b} S.-M. Fei \textit{et al.} (2009), {\it Experimental determination of entanglement for arbitrary pure states}, Phys. Rev. A, 80, 032320.

\bibitem{22} P. Horodecki (2003), {\it Measuring quantum entanglement without prior state reconstruction}, Phys. Rev. Lett., 90, 167901; P. Horodecki and A. Ekert (2002), {\it Method for direct detection of quantum entanglement}, Phys. Rev. Lett., 89, 127902.

\bibitem{23} H. A. Carteret (2005), {\it Noiseless quantum circuits for the Peres separability criterion}, Phys. Rev. Lett., 94, 040502; H. A. Carteret (2006), {\it Exact interferometers for the concurrence and residual 3-tangle}, arXiv:quant-ph/0309212; 
Y.-K. Bai \textit{et al.} (2006), {\it Method for detecting without the structural physical approximation by local operations and classical communication}, J. Phys. A: Math. Gen., 39, pp. 5847-5856.

\bibitem{24} R. Augusiak \textit{et al.} (2008), {\it Universal observable detecting all two-qubit entanglement and determinant-based separability tests}, Phys. Rev. A, 77, 030301(R).

\bibitem{18} J. Cai and W. Song (2008), {\it Novel schemes for directly measuring entanglement of general states}, Phys. Rev. Lett., 101, 190503.

\bibitem{20} A. Salles \textit{et al.} (2006), {\it Single observable concurrence measurement without simultaneous copies}, Phys. Rev. A, 74, 060303(R).


\bibitem{30} F. Mintert \textit{et al.} (2005), {\it Measures and dynamics of entangled states}, Phys. Rep. 415, pp. 207-259.

\bibitem{31} A. Borras \textit{et al.} (2009), {\it Typical features of the Mintert-Buchleitner lower bound for concurrence}, Phys. Rev. A, 79, 022112.

\bibitem{32} F. Mintert \textit{et al.} (2004), {\it Concurrence of mixed bipartite quantum states in arbitrary dimensions}, Phys. Rev. Lett., 92, 167902.

\bibitem{33} P. Rungta and C. M. Caves (2003), {\it Concurrence-based entanglement measures for isotropic states}, Phys. Rev. A, 67, 012307.

\bibitem{34} M. B. Plenio and S. Virmani (2007), {\it An introduction to entanglement measures}, Quant. Inform. Comput., 7, pp. 1-51.

\bibitem{34a} W. K. Wootters (1998), {\it Entanglement of formation of an arbitrary state of two qubits}, Phys. Rev. Lett., 80, pp. 2245-2248.

\bibitem{35} K. Chen \textit{et al.} (2005), {\it Concurrence of arbitrary dimensional bipartite quantum states}, Phys. Rev. Lett., 95, 040504.

\bibitem{36} X.-H. Gao \textit{et al.} (2006), {\it Lower bounds of concurrence for tripartite quantum systems}, Phys. Rev. A, 74, 050303(R).

\bibitem{36a} L. Li-Guo \textit{et al.} (2009), {\it A lower bound on concurrence}, Chin. Phys. Lett., 26, 060306.

\bibitem{36b} O. Gittsovich and O. Guhne (2010), {\it Quantifying entanglement with covariance matrices}, Phys. Rev. A, 81, 032333.

\bibitem{37} C.-. Yu \textit{et al.} (2008), {\it Evolution of entanglement for quantum mixed states}, Phys. Rev. A, 78, 062330.

\bibitem{38} Y.-C. Ou \textit{et al.} (2008), {\it Proper monogamy inequality for arbitrary pure quantum states}, Phys. Rev. A, 78, 012311.

\bibitem{39} M. Li \textit{et al.} (2009), {\it A lower bound of concurrence for multipartite quantum states}, J. Phys. A: Math. Theor., 42, 145303.

\bibitem{40} S. J. Akhtarshenas (2005), {\it Concurrence vectors in arbitrary multipartite quantum systems}, J. Phys. A: Math. Gen., 38, pp. 6777-6784.

\bibitem{41} M. Horodecki \textit{et al.} (1997), {\it Inseparable two spin-1/2 density matrices can be distilled to a singlet form}, Phys. Rev. Lett., 78, pp. 574-577;
M. Horodecki \textit{et al.} (1998), {\it Mixed-state entanglement and distillation: Is there a "bound" entanglement in nature?}, Phys. Rev. Lett., 80, pp. 5239-5242.

\bibitem{42} S. J. van Enk (2006), {\it Can measuring entanglement be easy?}, arXiv:quant-ph/0606017; S. J. van Enk \textit{et al.} (2007), {\it Experimental procedures for entanglement verification}, Phys. Rev. A, 75, 052318; S. J. van Enk (2009), {\it Direct measurements of entanglement and permutation symmetry}, Phys. Rev. Lett., 102, 190503.

\bibitem{42b} R. Demkowicz-Dobrzanski \textit{et al.} (2006), {\it Evaluable multipartite entanglement measures: Multipartite concurrences as entanglement monotones}, Phys. Rev. A, 74, 052303.

\bibitem{43} G. Vidal and R. F. Werner (2002), {\it Computable measure of entanglement}, Phys. Rev. A, 65, 032314.

\end{thebibliography}
\end{document}